\documentclass[12pt]{article}
\title{Epistemological-Scientific Realism and the Onto-Relationship of Inferentially Justified and Non-Inferentially Justified Beliefs}
\author{Max L. E. Andrews\footnote{mlandrews@liberty.edu}\\
\emph{Department of Philosophy, Liberty University}}
\date{13 May 2012}

\begin {document}
\maketitle

\begin{abstract}
The traditional concept of knowledge is a justified true belief.  The bulk of contemporary epistemology has focused primarily on that task of justification.  Truth seems to be a quite obvious criterionÑdoes the belief in question correspond to reality?  My contention is that the aspect of ontology is far too separated from epistemology.  This onto-relationship between reality and beliefs require the epistemic method of epistemological realism.  This is not to diminish the task of justification.  I will then discuss the role of inference from the onto-relationships of free invention and discovery and whether it is best suited for a foundationalist or coherentist model within a theistic context.\footnotemark[1]\footnotetext[1]{For inference to the best explanation (IBE) to be reliable, there must be a positive correlation between explantoriness and truth.  As discussed by Igor Douven in Ò"Testing Inference to the Best Explanation,"Ó \emph{Synthese} 130 (2002), IBE really own works with a realist conception.  If IBE is to be accepted as a legitimate methodology it must be substantiated in a self-affirming way, empirically evidenced, or privileged.  Douven argued that IBE is a testable methodology in a self-affirming way.  If that does not work then it would have to be a privileged belief (or basic belief) since it is certainly not empirically justifiable.}
\\
\\
\\
\end{abstract}

\section{The Ontological Status of Reality and its Relationship to Epistemology}

Should it be the case that metaphysical naturalism be false (including different modifications of simple or neo-Platonism) and theism be true it seems to be the case that God created our cognitive faculties in such a way that there is a certain fit or match between such faculties and the world.  This is called the adequation of the intellect to reality (\emph{adequation intellectus ad rem}).  The main premise to this adequation is that there is an onto-relationship between our cognitive or intellectual faculties and reality that enables us to know something about the world, God, and ourselves.\footnotemark[2]\footnotetext[2]{Alvin Plantinga, \emph{Where the Conflict Really Lies: Science, Religion, and Naturalism} (Oxford: Oxford University Press, 2011), 269.} This immanent rationality inherent to reality is not God but it does cry aloud for God if only because the immanent rationality in nature does not provide us with any explanation of itself.\footnotemark[3]\footnotetext[3]{John Morrison, \emph{Knowledge of the Self-Revealing God in the Thought of Thomas Forsyth Torrance} (Eugene, OR: Wipf and Stock, 1997), 106. Thomas Torrance, \emph{God and Rationality} (Oxford: Oxford University Press, 1971), 93-94.}

In reality all entities are ontologically connected or interrelated in the field in which they are found.  If this is true then the relation is the most significant thing to know regarding an object.  Thus, to know entities as they actually are what they are in their relation ÒwebsÓ.  Thomas Torrance termed this as onto-relations, which points more to the entity or reality, as it is what it is as a result of its constitutive relations.\footnotemark[4]\footnotetext[4]{Morrison, 106.}

The methodology of the epistemological realist concerns propositions of which are \emph{a posteriori}, or Òthinking after,Ó the objective disclosure of reality.  Thus, epistemology follows from ontology.  False thinking or methodology (particularly in scientific knowledge) has brought about a failure to recognize the intelligibility actually present in nature and the kinship in the human knowing capacity to the objective rationality to be known.\footnotemark[5]\footnotetext[5]{Thomas Torrance, \emph{Theological Science} (Oxford: Oxford University Press, 1969), 76-80.}

Lorenzo Valla (1406-1457) developed the interrogative (\emph{interrogatio}) rather than the problematic (\emph{quaestio}) form of inquiry.  VallaÕ's mode of inquiry was one in which questions yield results that are entirely new, giving rise to knowledge that cannot be derived by an inferential process from what was already known.  This method was similar to the works of Stoic lawyers and educators like Cicero and Quintilian; that is, questioning witnesses, investigating documents and states of affairs without any prior conception of what the truth might be.  Valla transitioned from not only using this method for historical knowledge but also applied it as Òlogic for scientific discovery.Ó\footnotemark[6]\footnotetext[6]{Thomas Torrance, Ò"Einstein and Scientific Theology,Ó" \emph{Religious Studies} 8 no. 3 (1972): 236-237.}

VallaÕ's logic for discovery was the art of finding out things rather than merely the art of drawing distinctions and connecting them together.  He called for an active inquiry (\emph{activa inquisitio}).  John Calvin (1509-1564) applied this method to the interpretation of Scripture and thus became the father of modern biblical exegesis and interpretation.\footnotemark[7]\footnotetext[7]{Valla served in conjunction with Andrea Alciati (1492-1550) as CalvinÕ's primary influence for his biblical interpretation.}  Francis Bacon (1561-1626) applied it to the interpretation of the books of nature, as well as to the books of God, and became the father of modern empirical science.\footnotemark[8]\footnotetext[8]{Torrance, "ÒEinstein,Ó" 237.}

This methodology created a split between subject and object, knowing and being, and gave rise to phenomenalism.  Newton claimed that he invented no hypotheses but deduced them from observations produced rationalistic positivism, which engulfed contemporary European thought.  This splitÕs gulf was widened by David Hume'Õs (1711-1776) criticism of causality and inference, depriving knowledge of any valid foundation in necessary connections obtaining between actual events and of leaving it with nothing more reliable than habits of mind rooted in association.\footnotemark[9]\footnotetext[9]{Ibid., 240.}   Hume weighed heavy in Immanuel Kant'Õs (1724-1804) philosophical development.  Given the Newtonian understanding of space and time, Kant transferred absolute space and time from the divine sensorium to the mind of man (the transfer of the inertial system), thus intellect does not draw its laws out of nature but imposes its laws upon nature.  According to Kant one cannot know the \emph{Ding an Sich} (thing itself) by pure reason; one is therefore limited to the sensual and shaping mental categories of the mind.  That which comes through sensation the intuitions are shaped by the mindÕ's \emph{a priori} categories.  It is in this sense that Kant played an essential part in the development of the idea that man is himself the creator of the scientific world.

Throughout Albert EinsteinÕ's work, the mechanistic universe proved unsatisfactory.  This was made evident after the discovery of the electromagnetic field and the failure of Newtonian physics to account for it in mechanistic concepts.  Then came the discovery of four-dimensional geometry and with it the realization that the geometrical structures of Newtonian physics could not be detached from changes in space and time with which field theory operated.  Einstein stepped back into stride with Newton and his cognitive instrument of free invention.  It was free in the sense that conclusions were not reached under logical control from fixed premises, and it was invented under the pressure of the nature of the universe upon the intuitive apprehension of it.  Einstein used Newton and MaxwellÕ's partial differential equations in field theory to develop a mode of rationality called mathematical invariance.  Mathematical invariance established a genuine ontology in which the subject grips with objective structures and intrinsic intelligibility of the universe.\footnotemark[10]\footnotetext[10]{Ibid., 241-42.}

This also meant a rejection of KantÕ's synthetic \emph{a priori} whereby knowledge of the phenomenal world is said to be reduced to an ÒorderÓ without actual penetration into the \emph{Ding an Sich}.\footnotemark[11]\footnotetext[11]{Morrison, 90.}  Einstein'Õs categories are not some form of Kantian \emph{a priori} but conceptions that are freely invented and are to be judged by their usefulness, their ability to advance the intelligibility of the world, which is dependent of the observer.  As he sees it, the difference between his own thinking and KantÕ's is on just this point:  Einstein understands the categories as free inventions rather than as unalterable (conditioned by the nature of the understanding).\footnotemark[12]\footnotetext[12]{Donna Teevan, Ò"Albert Einstein and Bernard Lonergan on Empirical Method,Ó" \emph{Zygon} 37 no. 4 (2002): 875-876.}   It is by this method that one can penetrate the inner rationality of the reality by discovery, imagination, and insight in order to construct forms of thought and knowledge through which the rationality of the object may be discerned.\footnotemark[13]\footnotetext[13]{Morrison, 105}   EinsteinÕ's free invention is quite synonymous with discovery in the sense that the consequent conclusion (knowledge) is not inferred or entailed from a fixed categorical antecedent (i.e. Kant).

Principles of method are closely related to empirical observations.  As Einstein put it, Òthe scientist has to worm these general principles out of nature by perceiving in comprehensive complexes of empirical facts certain general features which permit of precise formulation.Ó\footnotemark[14]\footnotetext[14]{Albert Einstein, \emph{Ideas and Opinions}, trans. and rev. Sonja Bargmann (New York: Three Rivers, 1982), 221.}  These principles, not Òisolated general laws abstracted from experienceÓ or Òseparate results from empirical research,Ó provide the basis of deductive reasoning.\footnotemark[15]\footnotetext[15]{Teevan, 877.}

\section{What About \emph{A Priori} and Non-Empirical Knowledge?}

The onto-relationships as described above concerning the intricate web and connection between reality and its entailment of knowledge does not seem to have such effect on \emph{a priori} and non-empirical knowledge.\footnotemark[16]\footnotetext[6]{To claim that such inferential reasons are not good reasons for belief one might deny the legitimacy of such forms of abductive reasoning as described above. The most common objection to such reasoning is when the conclusion of the argument involves unobservables (physically or metaphysically).  Stephen Leeds, Ò"Correspondence Truth and Scientific Realism,Ó" \emph{Synthese} 159 (2007): 3.  Bas van Fraassen takes the stronger objection to this inferential reasoning no matter what the context is; even if it is empirical \emph{a posteriori}.  For more on addressing van FraassenÕ's objection see Douven (1999).}   Such methodology inevitably turns all such knowledge into scientific knowledgeÑso what about ethical and religious knowledge\footnotemark[17]\footnotetext[17]{For the role of moral knowledge in non-inferential reasoning see Bart Streumer, Ò"Inferential and Non-Inferential Reasoning,Ó" \emph{Philosophy and Phenomenological Research} 74 (2007): 4-5.}?  Kant argued that such synthetic \emph{a priori} knowledge was logically prior to any \emph{a posteriori} knowledge.  Such knowledge would be excluded from inferential knowledge but not necessarily excluded form the onto-relationship with reality.  This knowledge may serve as an intuitive apprehension into the actual intrinsic relations in reality (physical and metaphysical).  This intuitive knowledge is rational but non-logical and non-inferential.  This could be said that it is the knowledge that serves as the foundations, which arise in the mindÕs assent under the impress of objective structures in reality.\footnotemark[18]\footnotetext[18]{Morrison, 91.}   There is no reason to limit such intuitive apprehension of reality to the physical world only, which would serve as a defeater for any further entailments for positivism or strict empiricism.  Such structures of reality may be purely metaphysical such as minds, abstract objects, or God.  However, there must be some type of causal capacity for the onto-relations to have effect, which would exclude abstract objects since they do not seem to stand in causal relations.  Thus, minds and God may serve as plausible ontological origins for non-empirical knowledge.

This methodology is not so far astray from the epistemological realistÕs empiricism, such a methodology I have assumed thus far, since the onto-relationship has still been preserved.  This form of method has replaced \emph{a posteriori} knowledge with \emph{a priori} but the apprehension of such knowledge is still preserved by the onto-relationship of reality.  Moral intuition may serve as an \emph{a priori} conception, which can be expressed either doxastically or in a self-evident or incorrigible way.  I do not see any good reason for why moral judgments should not function as evidence for a belief.  These judgments are not empirically based but intuitively based.  These intuitions are objective and are grounded in an objective reality, just as is any other criterion for evidence by empirical standards.  The only differentiation between moral intuitions and empirical judgments is whether they are \emph{a priori} or \emph{a posteriori} but are still harmonious with epistemological realism and the onto-relationship between reality and knowledge.  This causal relationship may simply be impressed upon us logically prior to our experience.\footnotemark[19]\footnotetext[19]{If the epistemological realistÕ's need for empiricism must be appeased by some experiential medium then it may certainly follow that the knowledge of certain ethical and religious truths may certainly come about \emph{a posteriori} as well, though this is not the typical approach or Ô'categoryÕ' for such knowledge.}

\section{Inferential Justification in Foundationalism and Coherentism}

Logically prior to such inferential reasoning is intuition for reasons previously discussed.  These intuitions may be basic beliefs. The belief that this glass of water in front of me will quench my thirst if I drink it is not inferred back from previous experiences coupled with an application of a synthetic \emph{a priori} principle of induction.  Though this example is not how we form our beliefs psychologically or historically, it can be formed via instances of past experience and induction in the logical sense.  However, when it does come to inferential reasoning R.A. Fumerton provides two definitions for what it means to say that one has inferential justification.\footnotemark[20]\footnotetext[20]{R.A. Fumerton, Ò"Inferential Justification and Empiricism,Ó" \emph{The Journal of Philosophy} 73 (1976): 564-65.}
\\
\\
\newpage

\noindent{D1 = S has an inferentially justified belief in P on the basis of E.}

(1) S believes P.

(2) S justifiably believes both E and the proposition that E confirms P.

(3) S believes P because he believes both E and the proposition that E confirms P.

(4) There is no proposition X such that S is justified in believing X and that E\&X does not confirm P.
\\
\\
D2 = S has an inferentially justified belief in P on the basis of E.

(1) S believes P.

(2) E confirms P.

(3) The fact that E causes S to believe P.

(4) There is no proposition X such that S is justified in believing X and that E\&X does not confirm P.
\\

Given the explications of such definitions, both D1 and D2, there seems to be good grounds for believing that P must be inferentially justified.  It is most certainly that case that D2 is more amenable to having scientific knowledge in the sense that both (2) and (3) are confirmatory.  D2-(3) is certainly difficult to substantiate without begging the question.  Having E cause S to believe P is difficult to distance from some form of transitive relation.  Inferential justification may also be expressed probabilistically or determined probabilistically.\footnotemark[21]\footnotetext[21]{This may be expressed by Thomas Bayes'Õ theorem for conditional probability: $p(H|E)=p(H)p(E|H)/p(E)$ or by his rule for belief change: $Pe(H) = p(H|E)$. If my belief p is going to be justified probabilistically then it must be greater than 0 and less than or equal to 1 where p is $> .5$.  Suppose that after all the evidence that is available is possessed and I have come to a value of precisely .5 for p.  If I reject p as being true then I have just as much of a chance of being wrong about that as I do as being right.  When p has a value of .5, all things considered, then I believe it would be acceptable to be believe p, ~p, or to be agnostic.  For more on the role of probability in inferential reasoning see Igor DouvenÕs Ò"Inference to the Best Explanation Made Coherent,Ó" \emph{Philosophy of Science} 66 (1999): S424-S435.} I have little contention with such definitions of inferential justification; my concern is whether this is most amicable within a foundationalistÕ's or coherentistÕ's noetic structure.

Both D1 and D2 offer, I believe, to be successful accounts of inferential justification.  However, I do find both definitions to be problematic for the empiricist on the bases of foundationalism, of which I will argue that such inferential justification and non-epistemological direct realism is more amicable to the coherentist and that a non-epistemological realist who adheres to foundationalism cannot successfully account for new beliefs.

Such inferential justification is certainly compatible with foundationalism but making all empirical claims to be inferential seems to be over-committing to inferential reasoning.  Suppose I am walking in the field and on the next hill over I see an object.  For all purposes, my phenomenological faculties indicate to me that there is something on the next hill.  This belief is held for a reason, primarily that my phenomenological faculties inform me that something is on the next hill over, but this is not a reasoned belief.  I may certainly infer certain properties consistent with D1 and D2 such as the belief that the object has a particular color or that it omits a certain sound or that it has a particular smell.  My belief that an object is on the next hill over from me seems to be quite basic.  I am not inferring its existence from other object-likenesses.  I am completely unaware as to the identity of this object, or better yet, whether this object is unique or unknown.  Suppose that this object has never been known before I experienced it.  This makes the situation quite different from FumertonÕ's glass of water and is not a future tensed proposition nor is it a subjunctive conditional.

Inferential reasoning as described by D1 and D2 are certainly kind to empiricism when it comes to scientific knowledge.  Certain unknown entities may become known by inferential means.  We can infer the existence of protons, quarks, and other elementary particles by predicting what effects such entities may have in certain situations.  This may be causal in nature and may be confirmed by inference. However, it is not the case that we directly experience the existence of these particles (for all intents and purposes, it certainly is the case that we experience particles when we run in to a wall and even then we experience the strong nuclear force over the particles). Nevertheless, epistemological direct realism and new belief formation can be non-inferentially justified.\footnotemark[22]\footnotetext[22]{This is not to ignore other experiential data such as religious experience.  Other propositional beliefs may be basic but non-empirical such as mathematical truths.  My concern is oriented towards empirical basic beliefs.  Additionally, suppose that today is Friday.  I cannot change my belief to believe that it is now Sunday or Monday.  Some beliefs are non-inferentially justified and involuntary. Richard Swinburne, \emph{Faith and Reason} (Oxford: Oxford University Press, 1981) 25.}

With such a methodology for inferential reasoning it may be argued that the foundationalist framework requires a presupposing of coherentism.  This would bring inference to the best explanation into close contact with the holistic view of explanation.\footnotemark[23]\footnotetext[23]{Philip Clayton, Ò"Inference to the Best Explanation,"Ó \emph{Zygon} 32 no. 3 (1997): 387.}  Philip Kitcher argued that this holistic view of inferential reasoning

\begin{quote}
[holds] that [scientific] understanding increases as we decrease the number of independent assumptions that are required to explain what goes on in the worldÉ Explanations serve to organize and systematize our knowledge in the most efficient and coherent possible fashion.  Understanding, on this view, involves having a world-pictureÑa scientific WeltanschauungÑand seeing how various aspects of the world and our experience of it fit into that picture.\footnotemark[24]\footnotetext[24]{Philip Kitcher, Ò"Scientific Explanation,"Ó \emph{Minnesota Studies in the Philosophy of Science} 13 (1989): 182.}
\end{quote}

Inferentially justified empirical beliefs are more in sync with a coherentist noetic structure.  When making inferential claims the proposition being inferred from must cohere to a proposition already accepted as truth.  Inferential reasoning is not necessarily non-foundational, but if empirical claims are strictly inferential then coherentism is best suited.  No matter what the belief in question is to be it must be inferentially referred back to another experientially valid belief (within the scope of empirical discussion).

David Hume brought to our attention a problem with inferential reasoning, which is especially important to the present task given his empiricism.

\begin{quote}
As to past experience, it can be allowed to give direct and certain information of those precise objects only and that precise period of time which fell under it cognizance.  But why this experience should be extended to future times, and to other objects, which for all we know, may be similar in appearanceÉThis, I must confess, seems to be the same difficultyÉThe question still recurs: on what process of argument this inference is founded?  Where is the medium, the interposing ideas which join propositions so very wide of each other?\footnotemark[25]\footnotetext[25]{David Hume, \emph{An Enquiry Concerning Human Understanding}, sect. IV, 2, in \emph{Philosophical Inquiry} Eds. Jonathan E. Adler and Catherine Z. Elgin (Indianapolis, IN: Hackett, 2007), 181-82.}
\end{quote}

Hume is correct, it does not follow.  There are plenty of possible worlds that match the actual world up to the present time, but then diverge wildly, so that inductive inferences would mostly fail in those other worlds.  It is by no means inevitable that inductive reasoning should be successful; its success is one more example of the fit between our cognitive faculties and the world.\footnotemark[26]\footnotetext[26]{Plantinga, 295-97.}   The criteria for the best inference are simplicity, beauty, and consilience (fit with other favored or established hypotheses).\footnotemark[27]\footnotetext[27]{These criteria may certainly be unnecessary in the case of paradigm shift with warranted evidence (preservation of consilience).  Additionally, beauty and simplicity are certainly preferred but as long as the inference is in relation to reality then these two criterions may be inapplicable.  Consilience is the most important criterion.}   Inferentially justified new beliefs create less dissonance with coherentism than with foundationalism.  What is needed logically prior to the acceptance or justification of new belief is an evidence base.  This is the set of beliefs used, or appealed to, in conducting an inquiry.\footnotemark[28]\footnotetext[28]{Ibid., 167.}   Recall TorranceÕ's onto-relations.  This onto-relation allows for inference to be a bridge between the ontological-epistemological divide.  It is the onto-relationship that serves as HumeÕs missing medium.  It is this Ò"webÓ" of onto-relations and consilience that function best with coherentism.  Thus, to think rightly and in terms of inference and \emph{a posteriori} reasoning means to connect things up with other things, thinking their constituent interrelations, and thus it is important for thinking to determine what kind of relation that exists between the realities contemplated.\footnotemark[29]\footnotetext[29]{Morrison, 107.}

\section{Conclusion}

There is a long historical development of onto-relations and inferential reasoning with primary influence by the contemporary science of the twentieth century and philosophers.  Inferential reasoning is a widely practiced methodology in the contemporary spheres of science, the philosophy of religion, and the philosophy of science.  Bas van Fraassen, as an antirealist, is one of the leading opponents of such inferential reasoning and its use of the probability calculus.  Despite Alvin Plantinga'Õs reliabilism he has made recent contributions to the onto-relations and scientific knowledge.\footnotemark[30]\footnotetext[30]{PlantingaÕ's reliabilism would serve as an antecedent to scientific knowledge just like foundationalism would as previously discussed.  However, it is difficult to separate the external element from the internalist nature of the task.} As Robert Audi put it, the contemporary task is discerning whether inferential and scientific knowledge is best suited for foundationalism or coherentism.\footnotemark[31]\footnotetext[31]{Robert Audi, \emph{Epistemology} ed. 3, (New York: Routledge, 2011), 300-1.}

This \emph{a posteriori} methodology inevitably turns all such knowledge into scientific knowledge.  Despite all \emph{a posteriori} knowledge being scientific in nature the onto-relations are preserved in emph{a priori} non-empirical knowledge.  Intuitions and basic beliefs may serve as antecedents for further inductive reasoning from which to use as the evidence base for the ÒwebÓ of consilience and onto-relations.\footnotemark[32]\footnotetext[32]{This goes to show that one belief would require the logically antecedent justification of one or more other beliefs.  This raises the problem of regress but if a basic belief serves as the unjustified justifier for that belief then it may be a justified antecedent.  Lawrence Bonjour, Ò"The Coherence Theory of Empirical Knowledge,"Ó in Paul Moser, \emph{Empirical Knowledge} (Totowa, NJ: Rowman \& Littlefield, 1986), 117.} Such methodological and inferential reasoning is not necessarily restricted to foundationalism, as Fumerton had argued since there are legitimate onto-relational basic beliefs.  If these onto-relational beliefs serve as antecedents from which further inductive or abductive reasoning is used then inferential reasoning becomes better understood when it is justified by other doxastic elements in the onto-relational Ò"web"Ó.

\end{document}